\newif\ifpreprint

\preprintfalse 

\ifpreprint
\documentclass[journal=jctcce,manuscript=letter]{achemso}
\else
\documentclass[journal=jctcce,manuscript=letter,layout=twocolumn]{achemso}
\fi

\usepackage[T1]{fontenc} 

\usepackage{amsmath}
\usepackage{newtxtext,newtxmath}

\usepackage{graphicx}
\usepackage{dcolumn}
\usepackage{braket}
\usepackage{multirow}
\usepackage{threeparttable}
\usepackage{xspace}
\usepackage{verbatim}
\usepackage[version=4]{mhchem} 
\usepackage{comment}
\usepackage{color,soul}

\usepackage{mathtools}
\usepackage[dvipsnames]{xcolor}
\usepackage{xspace}
\usepackage{ifthen}

\usepackage{qcircuit}

\usepackage{graphicx,longtable,dcolumn,mhchem}
\usepackage{rotating,color}
\usepackage{lscape}
\usepackage{amsmath}
\usepackage{dsfont}
\usepackage{soul}
\usepackage{physics}
\newcolumntype{d}{D{.}{.}{-1}}

\newcommand{\ie}{\textit{i.e}}
\newcommand{\eg}{\textit{e.g}}

\newcommand{\Pop}{6-31+G(d)}
\newcommand{\AVDZ}{\emph{aug}-cc-pVDZ}
\newcommand{\AVTZ}{\emph{aug}-cc-pVTZ}
\newcommand{\AVQZ}{\emph{aug}-cc-pVQZ}
\newcommand{\AVFZ}{\emph{aug}-cc-pV5Z}
\newcommand{\Td}{\%T_1}

\usepackage[colorlinks = true,
            linkcolor = blue,
            urlcolor  = black,
            citecolor = blue,
            anchorcolor = black]{hyperref}

\definecolor{goodorange}{RGB}{225,125,0}
\definecolor{goodgreen}{RGB}{5,130,5}
\definecolor{goodred}{RGB}{220,50,25}
\definecolor{goodblue}{RGB}{30,144,255}

\newcommand{\note}[2]{
\ifthenelse{\equal{#1}{F}}{
\colorbox{goodorange}{\textcolor{white}{\footnotesize \fontfamily{phv}\selectfont #1}}
    \textcolor{goodorange}{{\footnotesize \fontfamily{phv}\selectfont #2}}\xspace
}{}
\ifthenelse{\equal{#1}{R}}{
\colorbox{goodred}{\textcolor{white}{\footnotesize \fontfamily{phv}\selectfont #1}}
    \textcolor{goodred}{{\footnotesize \fontfamily{phv}\selectfont #2}}\xspace
}{}
\ifthenelse{\equal{#1}{N}}{
\colorbox{goodgreen}{\textcolor{white}{\footnotesize \fontfamily{phv}\selectfont #1}}
    \textcolor{goodgreen}{{\footnotesize \fontfamily{phv}\selectfont #2}}\xspace
}{}
\ifthenelse{\equal{#1}{M}}{
\colorbox{goodblue}{\textcolor{white}{\footnotesize \fontfamily{phv}\selectfont #1}}
    \textcolor{goodblue}{{\footnotesize \fontfamily{phv}\selectfont #2}}\xspace
}{}
}

\usepackage{titlesec}

\usepackage[fontsize=11pt]{scrextend}
\titleformat{\section}
{\normalfont\sffamily\bfseries\color{Blue}}
{\thesection.}{0.25em}{\uppercase}

\titleformat{\subsection}
{\normalfont\sffamily\bfseries}
{\thesubsection}{0.25em}{}

\titleformat{\subsubsection}
{\normalfont\sffamily}
{\thesubsubsection}{0.25em}{}

\titleformat{\suppinfo}
{\normalfont\sffamily\bfseries}
{\thesubsection}{0.25em}{}

\titlespacing*{\section}{0pt}{0.5\baselineskip}{0.01\baselineskip}
\titlespacing*{\subsection}{0pt}{0.125\baselineskip}{0.01\baselineskip}
\titlespacing*{\subsubsection}{0pt}{0.125\baselineskip}{0.01\baselineskip}

\author{Pierre-Fran\c{c}ois Loos}
	\email{loos@irsamc.ups-tlse.fr}
	\affiliation[LCPQ, Toulouse]{Laboratoire de Chimie et Physique Quantiques, Universit\'e de Toulouse, CNRS, UPS, France}
\author{Anthony Scemama}
	\affiliation[LCPQ, Toulouse]{Laboratoire de Chimie et Physique Quantiques, Universit\'e de Toulouse, CNRS, UPS, France}
\author{Martial Boggio-Pasqua}
	\affiliation[LCPQ, Toulouse]{Laboratoire de Chimie et Physique Quantiques, Universit\'e de Toulouse, CNRS, UPS, France}
\author{Denis Jacquemin}
	\email{Denis.Jacquemin@univ-nantes.fr}
	\affiliation[CEISAM, Nantes]{Universit\'e de Nantes, CNRS,  CEISAM UMR 6230, F-44000 Nantes, France}
		
\let\oldmaketitle\maketitle
\let\maketitle\relax

     \title{A Mountaineering Strategy to Excited States: Highly-Accurate Energies and Benchmarks for Exotic Molecules and Radicals}

\date{\today}

\begin{tocentry}
	\centering
	\includegraphics[width=.8\textwidth]{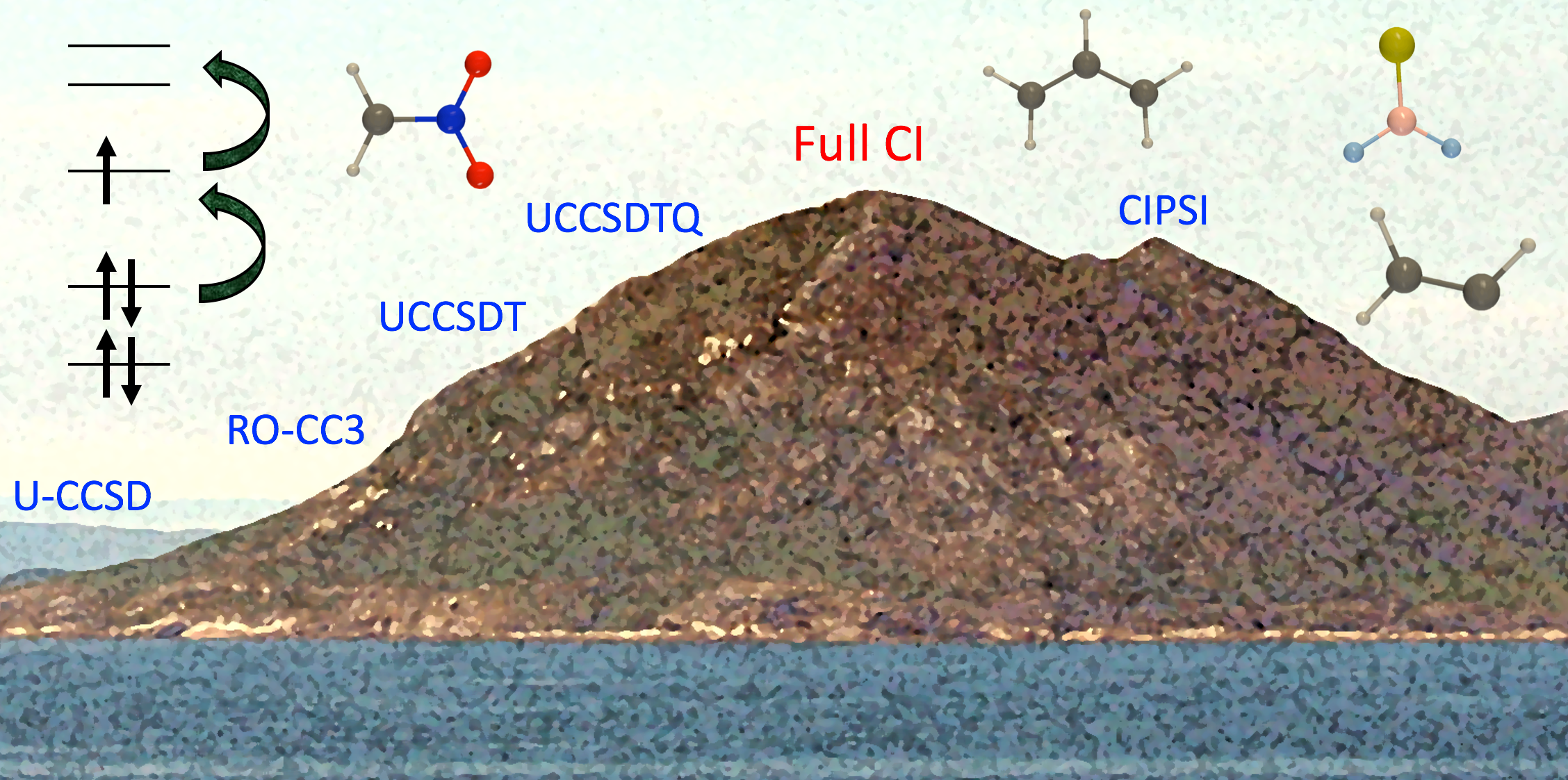}
\end{tocentry}

\begin{document}

\ifpreprint
\else
\twocolumn[
\begin{@twocolumnfalse}
\fi
\oldmaketitle

\begin{abstract}
Aiming at completing the sets of FCI-quality transition energies that we recently developed (\textit{J.~Chem.~Theory Comput.} \textbf{14} (2018) 4360--4379,  \textit{ibid.}~\textbf{15} (2019) 1939--1956, and \textit{ibid.}~\textbf{16} (2020) 
1711--1741), we provide, in the present contribution, ultra-accurate vertical excitation energies for a series of ``exotic'' closed-shell molecules containing F, Cl, P, and Si atoms and small radicals, such as CON and its variants, 
that were not considered to date in such investigations. This represents a total of 81 high-quality transitions obtained with a series of diffuse-containing basis sets of various sizes. For the exotic compounds, these transitions are used to 
perform benchmarks with a vast array of lower-level models, \textit{i.e.}, CIS(D), EOM-MP2, (SOS/SCS)-CC2, STEOM-CCSD, CCSD, CCSDR(3), CCSDT-3, (SOS-)ADC(2), and ADC(3). Additional comparisons are made with literature data. 
For the open-shell compounds, we have compared the performances of both the unrestricted and restricted open-shell CCSD and CC3 formalisms.
\end{abstract}

\ifpreprint
\else
\end{@twocolumnfalse}
]
\fi

\ifpreprint
\else
\small
\fi

\noindent

\section{Introduction}

The increase of computational ressources coupled to the emergence of more advanced algorithms has led to a resurgence of the selected configuration interaction (SCI) approaches \cite{Ben69,Whi69,Hur73} as an effective strategy to rapidly 
reach the full CI (FCI) limit at a fraction of the cost of a genuine FCI calculation thanks to a sparse exploration of the FCI space.\cite{Gin13,Caf14,Eva14,Gin15,Gar17b,Caf16,Caf16b,Sch16,Hol16,Liu16b,Sha17,Hol17,Chi18,Gar18,Sce18,Gar19} 
This revival is especially beneficial for the calculation of transition energies between electronic states, \cite{Eva14,Hol17,Gar18,Chi18,Sce18,Gar19,Loo18a,Loo19c,Gin19,Loo20a} as the accurate determination of these energies remains one of 
the great challenges faced by theoretical chemists. 

Recently, we have developed two sets of theoretical best estimates (TBEs) of FCI quality for the vertical transition energies of small closed-shell compounds. \cite{Loo18a,Loo19c} (See Ref.~\citenum{Loo20c} for a recent review.) 
In our first work, \cite{Loo18a} we reported TBEs for more than 100 electronic transitions of single-excitation character in organic compounds containing from one to three non-hydrogen atoms, namely \ce{C}, \ce{N}, \ce{O}, and \ce{S}. 
These TBEs have been obtained thanks to an efficient implementation of the CIPSI (Configuration Interaction using a Perturbative Selection made Iteratively) SCI algorithm, \cite{Gar19} which selects the most important determinants 
in the FCI space using a second-order perturbative criterion. \cite{Gar17b} Their quality was further confirmed by equation-of-motion coupled-cluster (CC) calculations performed up to high excitation degrees. It turned out that CC 
including contributions up to the quadruples (CCSDTQ) \cite{Kuc91} yields transition energies almost systematically equal to FCI, with a mean absolue error (MAE) as small as $0.01$ eV, whereas the three tested CC approaches 
including perturbative triples, namely, CC3, \cite{Chr95b,Koc97} CCSDT-3, \cite{Wat96,Pro10} and CCSDT \cite{Nog87} are also very effective with MAEs of $0.03$ eV. \cite{Loo18a}  This means that these four CC models are 
(on average) chemically accurate (error smaller than $1$ kcal.mol$^{-1}$ or $0.043$ eV) for these single excitation transitions. 

Our second set encompasses 20 transitions characterized by a large and/or dominant double excitation nature. \cite{Loo19c} These types of electronic excitations are known to be much more challenging for single-reference methods.  
For this set, we relied again on SCI methods to determine TBEs and we evaluated the performances of various multi-reference approaches, such as the second-order complete active space perturbation theory (CASPT2), \cite{Roo96,And90} 
and the second-order $n$-electron valence state perturbation theory (NEVPT2) \cite{Ang01,Ang01b,Ang02} methods.  Interestingly, for excitations with a large but not dominant double excitation character, such as the first $^1A_g$ 
excited state of \textit{trans}-butadiene, it turns out that the accuracy obtained with CC3 and NEVPT2 are rather similar with MAEs of ca.~$0.12$ eV.  \cite{Loo19c}  In contrast, for genuine double excitations (\ie, excitations with an insignificant 
amount of single excitation character) in which one photon effectively promotes two electrons, the CC3 error becomes extremely large (of the order of $1$ eV) and multi-reference approaches have clearly the edge (for example, the MAE of 
NEVPT2 is $0.07$ eV). \cite{Loo19c}  

To the very best of our knowledge, these two sets taken together constitute the largest ensemble of chemically-accurate vertical transition energies published to date with roughly $130$ transition energies of FCI quality.  Despite their 
decent sizes and the consideration of both valence and Rydberg excited states, these sets have obvious limitations. Let us point out four of these biases: (i) only small compounds are included; (ii) some important classes of transitions, 
such as charge-transfer (CT) excitations, are absent; (iii) compounds including only \ce{C}, \ce{N}, \ce{O}, \ce{S}, and \ce{H} atoms have been considered; (iv) these sets include only singlet-singlet and singlet-triplet 
excitations in closed-shell molecules.  

Very recently, we have made extensive efforts in order to solve the first limitation. \cite{Loo20a} However, performing SCI or high-level CC calculations rapidly becomes extremely tedious when one increases the system size as one 
hits the exponential wall inherently linked to these methods. At this stage, we believe that circumventing the second limitation is beyond reach as clear intramolecular CT transitions only occur in (very) large molecules for which CCSDTQ 
or SCI calculations remain clearly out of reach with current technologies.  We note, however, that intermolecular CT energies were recently obtained at the CCSDT level by Kozma and coworkers.  \cite{Koz20} Therefore, the aim of the present 
contribution is to get rid of the two latter biases. To this end, we consider here (i) a series of closed-shell compounds including (at least) one of the following atoms: \ce{F}, \ce{Cl}, \ce{Si}, or \ce{P}; (ii) a series of radicals characterized by 
open-shell electronic configurations and an unpaired electron. For the sake of simplicity, we denote the first additional set as ``exotic'' because it includes a series of chemical species that are rather unusual for organic chemistry, \eg, \ce{H-P=S} 
and \ce{H2C=Si}. Similar compounds were included in a benchmark set by the Ortiz group. \cite{Hah14} They were, however, using experimental data as reference, which often precludes straightforward comparisons with theoretical vertical transition
energies. \cite{Loo18b,Loo19b} On the other hand, the second set, simply labeled as ``radical'', encompasses doublet-doublet transitions in radicals. We believe that the additional FCI-quality estimates that we provide in the present study 
for both types of compounds nicely complete our previous works and will be valuable for the electronic structure community.

\section{Computational methods}

Our computational protocol closely follows the one of Ref.~\citenum{Loo18a}. Consequently, we only report key elements below. We refer the reader to our previous work for further information about the methodology and 
the technical details. \cite{Loo18a}
In the following, we report several statistical indicators: the mean signed error (MSE), mean absolute error (MAE), root-mean square error (RMSE), and standard deviation of the errors (SDE).

\subsection{Geometries and basis sets}

For the exotic set, we use CC3/{\AVTZ} ground-state geometries obtained without frozen-core (FC) approximation (\ie, correlating all electrons) to be consistent with our previously-published geometries. 
\cite{Bud17,Jac18a,Bre18a,Loo18a}  These optimizations have been performed using DALTON 2017 \cite{dalton} and CFOUR 2.1, \cite{cfour} applying default parameters. For the open-shell derivatives, the geometries 
are optimized at the UCCSD(T)/{\AVTZ} level using the GAUSSIAN16 program \cite{Gaussian16}  and applying the \textsc{tight} convergence threshold. The Cartesian coordinates of each compound are available in the 
Supporting Information (SI).

Throughout this paper, we use either the diffuse-containing Pople  {\Pop} basis set, or the Dunning \emph{aug}-cc-pVXZ (X $=$ D, T, Q, and 5) correlation-consistent family of atomic bases. 

\subsection{CC reference calculations}

The CC calculations are performed with several codes. For closed-shell molecules, CC3 \cite{Chr95b,Koc97} calculations are achieved with DALTON \cite{dalton} and CFOUR; \cite{cfour}  CCSDT calculations are performed 
with CFOUR \cite{cfour} and MRCC 2017;\cite{Rol13,mrcc} the latter code being also used for CCSDTQ and CCSDTQP.   Note that all our excited-state CC calculations are performed within the equation-of-motion (EOM) 
or linear-response (LR) formalism that yield equivalent excited-state energies. The reported oscillator strengths have been computed in the LR-CC3 formalism only. For open-shell molecules, the CCSDT, CCSDTQ, and 
CCSDTQP calculations performed with MRCC \cite{Rol13,mrcc} do consider an unrestricted Hartree-Fock (UHF) wave function as reference. All excited-state calculations are performed, except when explicitly mentioned, in 
the FC approximation using large cores for the third-row atoms. All electrons are correlated for the \ce{Be} atom, for which we systematically applied the basis set as included in MRCC. \cite{Pra10} (We have noted 
differences in the definition of the Dunning bases for this particular atom depending on the software that one considers.) 

\subsection{Selected Configuration Interaction}

All the SCI calculations are performed within the FC approximation using QUANTUM PACKAGE \cite{Gar19} where the CIPSI algorithm \cite{Hur73} is implemented. Details regarding this specific CIPSI implementation 
can be found in Refs.~\citenum{Gar19} and \citenum{Sce19}. We use a state-averaged formalism which means that the ground and excited states are described with the same number and same set of determinants, but 
different CI coefficients. The SCI energy is defined as the sum of the variational energy (computed via diagonalization of the CI matrix in the reference space) and a second-order perturbative correction which estimates the 
contribution of the determinants not included in the CI space. \cite{Gar17b}  By extrapolating this second-order correction to zero, one can efficiently estimate the FCI limit for the total energies, and hence, compute the 
corresponding transition energies. We estimate the extrapolation error by the  difference between the transition energies obtained with the largest SCI wave function and the FCI extrapolated value. These errors are 
systematically reported in the Tables below. Although this cannot be viewed as a true error bar, it provides a rough idea of the quality of the FCI extrapolation and estimate. 

\subsection{Other wave function calculations}

Our benchmark effort consists in evaluating the accuracy of vertical transition energies obtained at lower levels of theory. These calculations are performed with a variety of codes. For the exotic set, we 
rely on: GAUSSIAN  \cite{Gaussian16} and TURBOMOLE 7.3 \cite{Turbomole} for CIS(D); \cite{Hea94,Hea95} Q-CHEM 5.2 \cite{Kry13} for EOM-MP2 [CCSD(2)] \cite{Sta95c} and ADC(3); \cite{Tro02,Har14,Dre15} 
Q-CHEM  \cite{Kry13} and TURBOMOLE  \cite{Turbomole} for ADC(2); \cite{Tro97,Dre15} DALTON  \cite{dalton} and TURBOMOLE  \cite{Turbomole}  for CC2; \cite{Chr95,Hat00}  DALTON  \cite{dalton} and GAUSSIAN 
for CCSD;\cite{Pur82}  DALTON  \cite{dalton} for  CCSDR(3); \cite{Chr96b} CFOUR  \cite{cfour} for  CCSDT-3; \cite{Wat96,Pro10} and ORCA \cite{Nee12} for  similarity-transformed EOM-CCSD (STEOM-CCSD). \cite{Noo97,Dut18} 
In addition, we evaluate the spin-opposite scaling (SOS) variants of ADC(2), SOS-ADC(2), as implemented in both Q-CHEM, \cite{Kra13} and TURBOMOLE. \cite{Hel08} Note that these two codes have distinct SOS 
implementations, as explained in Ref.~\citenum{Kra13}. We also test the SOS and spin-component scaled (SCS) versions of CC2, as implemented in TURBOMOLE. \cite{Hel08,Turbomole} Discussion of various spin-scaling 
schemes can  be found elsewhere. \cite{Goe10a} When available, we take advantage of the resolution-of-the-identity (RI) approximation in TURBOMOLE and Q-CHEM. For the STEOM-CCSD calculations, it was checked that the 
active character percentage was, at least, $98\%$. When comparisons between various codes/implementations were possible, we could not detect variations in the transition energies larger than $0.01$ eV. For the radical set
molecules, we applied both the U (unrestricted) and RO (restricted open-shell) versions of CCSD and CC3 as implemented in the PSI4 code, \cite{Psi4} to perform our benchmarks.

\section{Results and Discussion}

\subsection{Exotic set}

\subsubsection{Reference values and comparison to literature}

\begin{table*}[htp]
\scriptsize
\caption{Excitation energies (in eV) of the exotic set obtained within the FC approximation. For each transition, we also report, on the left hand side, the LR-CC3/{\AVTZ} oscillator strength, the CC3 single excitation 
character ($\Td)$, and the TBE/{\AVTZ} excitation energy. Except otherwise stated, the latter has been obtained directly at FCI/{\AVTZ} level.  We also provide the TBE/CBS estimate obtained by correcting the 
TBE/{\AVTZ} value by the difference between CC3/\emph{aug}-cc-pV5Z and CC3/{\AVTZ}. On the right hand side, one finds the transition energies computed at various levels of theory.  T, TQ, and TQP 
stand for CCSDT, CCSDTQ, and CCSDTQP, respectively.} 
\label{Table-1}
\vspace{-0.3 cm}
\begin{tabular}{llp{1.0cm}p{.3cm}p{.5cm}p{.5cm}|p{.4cm}p{.4cm}p{.4cm}p{.4cm}|p{.4cm}p{.4cm}p{.4cm}p{1.1cm}|p{.4cm}p{.4cm}p{.4cm}p{1.1cm}}
\hline 
			&	&\multicolumn{3}{c}{\AVTZ}  & CBS & \multicolumn{4}{c}{\Pop} & \multicolumn{4}{c}{\AVDZ}  & \multicolumn{4}{c}{\AVTZ} \\
			&	& 	$f$ [CC3] &$\Td$ & TBE	& TBE			&CC3	& T		& TQ 	& TQP 	& CC3	&  T		& TQ 	& FCI 			&CC3	& T 		& TQ & FCI \\
\hline
Carbonylfluoride& $^1A_2$	&		&91.1&7.31$^a$	&7.31	&7.33	&7.30	&		&		&7.34	&7.31	&		&7.30$\pm$0.04	&7.31	&7.28	&		&7.32$\pm$0.05\\
			& $^3A_2$	&		&97.8&7.06$^b$	&7.07	&7.03	&7.00	&		&		&7.05	&7.02	&		&7.08$\pm$0.01	&7.03	&7.00	&		&7.04$\pm$0.10	\\
\ce{CCl2}		& $^1B_1$	&0.002	&93.7&2.59$^b$	&2.57	&2.71	&2.70	&2.70	&		&2.69	&2.69	&		&2.68$\pm$0.02	&2.61	&2.60	&		& \\
			& $^1A_2$	&		&88.3&4.40$^b$	&4.41	&4.46	&4.44	&4.47	&		&4.40	&4.39	&		&4.46$\pm$0.01 	&4.35	&4.33	&	 	&\\
			& $^3B_1$	&		&98.6&1.22$^b$	&1.23	&1.10	&1.09	&1.11	&		&1.20	&1.19	&		&1.22$\pm$0.03 	&1.20	&1.19	&		&1.22$\pm$0.05	 \\
			& $^3A_2$	&		&96.1&4.31$^b$	&4.32	&4.41	&4.38	&4.42	&		&4.34	&4.31	&		&4.36$\pm$0.01 	&4.28	&4.26	&		&				 \\		

\ce{CClF}		& $^1A''$		&0.007	&93.9&3.55$^b$	&3.54	&3.66	&3.66	&3.66	&		&3.63	&3.62	&		&3.62$\pm$0.01	&3.56	&3.55	&		&3.63$\pm$0.06	\\
\ce{CF2}		& $^1B_1$	&0.034	&94.7&5.09		&5.07	&5.18	&5.18	&5.18	&		&5.12	&5.11	&5.11	&5.12$\pm$0.00	&5.07	&5.06	&		&5.09$\pm$0.01	\\			
			& $^3B_1$	&		&99.1&2.77		&2.78	&2.71	&2.70	&2.71	&		&2.71	&2.70	&2.71	&2.71$\pm$0.01	&2.76	&2.75	&		&2.77$\pm$0.01	\\
Difluorodiazirine&$^1B_1$	&0.002	&93.1&3.74$^c$	&3.73$^d$&3.83	&3.83	&		&		&3.80	&3.80	&		&				&3.74	&3.74\\	
			&$^1A_2$		&		&91.4&7.00$^c$	&6.98$^d$&7.13	&7.11	&		&		&7.11	&7.08	&		&				&7.02	&7.00\\	
			&$^1B_2$		&0.026	&93.3&8.52$^c$	&8.54$^d$&8.51	&8.52	&		&		&8.45	&8.46	&		&				&8.50	&8.52\\	
			&$^3B_1$		&		&98.2&3.03$^e$	&3.03$^d$&3.09	&3.09	&		&		&3.06	&3.06	&		&				&3.03	&\\	
			&$^3B_2$		&		&98.9&5.44$^e$	&5.46$^d$&5.48	&5.48	&		&		&5.47	&5.46	&		&				&5.45	&\\	
			&$^3B_1$		&		&98.4&5.80$^e$	&5.81$^d$&5.86	&5.85	&		&		&5.83	&5.82	&		&				&5.81	&\\
Formylfluoride	& $^1A''$		&0.001	&91.2&5.96$^b$	&5.97	&6.09	&6.06	&6.07	&		&6.03	&6.00	&		&6.00$\pm$0.03	&5.99	&5.96	&		&\\
			& $^3A''$		&		&97.9&5.73$^b$	&5.75	&5.72	&5.70	&5.71	&		&5.65	&5.62	&		&5.65$\pm$0.01	&5.62	&5.60	&		&\\
\ce{HCCl}		& $^1A''$		&0.003	&94.5&1.98		&1.97	&2.05	&2.04	&2.05	&2.05	&2.02	&2.02	&2.02	&2.04$\pm$0.01	&1.97	&1.97	&		&1.98$\pm$0.00\\
\ce{HCF}		&$^1A''$		&0.006	&95.4&2.49		&2.49	&2.58	&2.57	&2.58	&2.58	&2.53	&2.53	&2.53	&2.54$\pm$0.00	&2.49	&2.49	&		&2.49$\pm$0.02\\
\ce{HCP}		& $^1\Sigma^-$	&		&94.9&4.84		&4.81	&5.19	&5.19	&5.18	&5.18	&5.06	&5.05	&5.04	&5.04$\pm$0.00	&4.85	&4.85	&4.84	&4.84$\pm$0.00\\
			& $^1\Delta$	&		&94.0&5.15		&5.10	&5.48	&5.48	&5.48	&5.48	&5.33	&5.33	&5.32	&5.32$\pm$0.00	&5.15	&5.15	&		&5.15$\pm$0.00\\
			& $^3\Sigma^+$&		&98.9&3.47		&3.49	&3.44	&3.45	&3.46	&3.46	&3.47	&3.47	&3.49	&3.49$\pm$0.00	&3.45	&3.45	&3.46	&3.47$\pm$0.00\\
			& $^3\Delta$	&		&98.8&4.22		&4.20	&4.40	&4.39	&4.39	&4.39	&4.35	&4.34	&4.34	&4.34$\pm$0.00	&4.22	&4.21	&4.21	&4.22$\pm$0.00\\
\ce{HPO}		& $^1A''$		&0.003	&90.9&2.47		&2.49	&2.49	&2.47	&2.48	&2.48	&2.47	&2.45	&2.46	&2.46$\pm$0.00	&2.46	&2.46	&		&2.47$\pm$0.00\\
\ce{HPS}		& $^1A''$		&0.001	&90.3&1.59		&1.61	&1.57	&1.55	&1.56	&1.56	&1.60	&1.59	&1.59	&1.60$\pm$0.00	&1.59	&1.58	&		&1.59$\pm$0.00\\
\ce{HSiF}		& $^1A''$		&0.024	&93.1&3.05		&3.05	&3.09	&3.08	&3.08	&3.08	&3.08	&3.07	&3.07	&3.06$\pm$0.00	&3.07	&3.06	&		&3.05$\pm$0.00\\
\ce{SiCl2}		&$^1B_1$		&0.031	&92.1&3.91$^b$	&3.93	&3.94	&3.94	&3.94	&		&3.93	&3.92	&		&3.95$\pm$0.02	&3.90	&3.88	&		&3.88$\pm$0.03\\
			&$^3B_1$		&		&98.7&2.48$^f$	&2.50	&2.39	&2.39	&2.40	&		&2.45	&2.44	&		&2.47$\pm$0.05	&2.48	&2.47	&		&2.49$\pm$0.04\\
Silylidene		&$^1A_2$		&		&92.3&2.11		&2.12	&2.14	&2.11	&2.10	&2.10	&2.18	&2.15	&2.14	&2.14$\pm$0.00	&2.15	&2.13	&2.12	&2.11$\pm$0.01\\
			&$^1B_2$		&0.033	&88.0&3.78		&3.80	&3.88	&3.87	&3.88	&3.88	&3.81	&3.80	&3.80	&3.79$\pm$0.01	&3.78	&3.78	&3.78	&3.78$\pm$0.01\\
\hline																			
\end{tabular}
\vspace{-0.3 cm}
\begin{flushleft}
$^a${FCI/{\Pop} value of 7.33$\pm$0.02 eV corrected by the difference between CCSDT/{\AVTZ} and CCSDT/{\Pop};}
$^b${FCI/{\AVDZ} value corrected by the difference between CCSDT/{\AVTZ} and CCSDT/{\AVDZ};}
$^c${CCSDT/{\AVTZ} value;}
$^d${Corrected with the quadruple-$\zeta$ basis rather than the quintuple-$\zeta$ basis;}
$^e${CCSDT/{\AVDZ} value corrected by the difference between CC3/{\AVTZ} and CC3/{\AVDZ};}
$^f${CCSDTQ/{\Pop} value corrected by the difference between CCSDT/{\AVTZ} and CCSDT/{\Pop}.}
\end{flushleft}
\end{table*}

Our main results are listed in Table \ref{Table-1} for the exotic set that encompasses 30 electronic transitions (19 singlets and 11 triplets) in 14 molecules containing between two and five non-hydrogen atoms.  Before briefly discussing the 
compounds individually, let us review some general trends. First, as one could expect for rather low-lying excitations, the {\AVTZ} basis set is sufficient large to provide excitation energies close to the complete basis set (CBS) limit \cite{Gin19} and the FC approximation 
is rather unimportant. Indeed, CC3 calculations performed with quadruple- and quintuple-$\zeta$ basis sets, with and without correlating the core electrons for the former basis, yield negligible changes as compared to the {\AVTZ} results. 
As more quantitatively illustrated by the results gathered in Table S1 of the SI, the maximal variation between CC3/{\AVTZ} and CC3/{\AVQZ} excitation energies is $0.03$ eV ($^1\Delta$ state of \ce{HCP}), and the MAE between the two 
basis sets is as small as $0.01$ eV. The same observation applies to the FC approximation with a mean absolute variation of $0.02$ eV between the CC3(full)/\emph{aug}-cc-pCVQZ and CC3(FC)/\emph{aug}-cc-pVQZ excitation energies. 
We therefore do not discuss further the quadruple- and quintuple-$\zeta$ results in the following, although basis set corrected TBEs can be found in Table \ref{Table-1}. Secondly, it can be seen, from the CC3 $\Td$ values (which provides 
a measure of the amount single excitation character of the considered transition) listed in Table \ref{Table-1}, that all the transitions considered here are largely dominated by single excitations, the smallest $\Td$ being $88\%$ (the 
second transition of silylidene). Such character is favorable to ensure a rapid convergence of the CC series. This is clearly exemplified by the convergence behavior of the {\Pop} excitation energies for which the CCSDTQ and the CCSDTQP 
transition energies are equal for the 11 cases for which the latter level of theory was achievable. Likewise, one notices that the CCSDTQ estimate systematically falls within $0.01$ eV of the FCI value that comes with a very small error bar for 
most transitions.  It is also reassuring to see that, for a given basis set, we could not detect variations larger than $0.04$ eV between CCSDTQ results and their CC3 and CCSDT counterparts, the changes being typically of ca.~$0.01$--$0.02$ 
eV. All these facts indicate that one can trust the FCI estimates, and hence the TBEs listed in Table \ref{Table-1} (for the larger difluorodiazirine molecule, see discussion below).

In the spirit of the famous Thiel paper, \cite{Sch08} let us now briefly discuss each compound and compare the results to available data. We do not intend here to provide an exhaustive review of previous calculations, which would lead to 
a gigantic list of references for the triatomic systems, but rather to pinpoint the ``best'' published excitation energies to date.

\emph{Carbonylfluoride.} For this compound encompassing four heavy atoms, the convergence of the SCI approach is rather slow and one notices a $0.03$ eV drop of the transition energies between CC3 and CCSDT. 
We therefore used FCI estimates determined with small bases, corrected for basis set effects to generate our TBEs.  For the lowest singlet, that is heavily blueshifted as compared to the parent formaldehyde, the most advanced previous 
theoretical studies reported vertical transition energies of $7.31$ eV [CCSDR(3)], \cite{Lav11} and $7.31$ eV [MRCI+Q]. \cite{Kat11} The measured EEL value is ca.~$7.3$ eV, \cite{Kat11} whereas the UV spectrum shows a peak at $7.34$ 
eV. \cite{Wor70}  All these values are obviously compatible with the current result.  Note that the interpretation of the measured 0-0 values for \ce{F2C=O} \cite{Jud83b} is challenging, as discussed elsewhere. \cite{Loo18b} For the triplet, 
the previous TBE is likely a $7.07$ eV MRCI+Q result, \cite{Kat11} also very close to our present value, whereas there also exists estimates of the triplet adiabatic energies. \cite{Bok09}

\emph{CCl$_2$, CClF, and CF$_2$.}  Dichlorocarbene is large enough to make the convergence of the SCI calculations difficult with the triple-$\zeta$ basis, and our TBEs are based on the FCI/\emph{aug}-cc-pVDZ values corrected for basis set effects
determined at the CC level. While both CC3 and CCSDT almost perfectly reproduce the FCI results for the singlet and triplet $B_1$ states, more significant differences are noted for the higher-lying $A_2$ states that seem slightly too low with 
CCSDT. This is also confirmed by the CCSDTQ results obtained with the Pople basis set.  Previous calculations are available at CCSD, \cite{Cze07} and MRCI \cite{Cai93,Sun15b} levels.  The most recent MRCI+Q values, obtained with a large atomic basis 
set are $2.61$, $4.49$, $1.25$ and $4.43$ eV for the $^1B_1$, $^1A_2$, $^3B_1$, and $^3A_2$ transitions, respectively. These values are reasonably close to the present TBEs. For CClF, the most accurate literature value is probably the 
MRCI+Q/triple-$\zeta$ estimate of $3.59$ eV, \cite{Sun15c} within $0.03$ eV of our current TBE. For this compound, we are also aware of three previous experimental investigations focussing on its vibronic spectra. \cite{Sch91,Kar93,Gus01} For \ce{CF2}, 
the SCI calculations converge rapidly even with the {\AVTZ} basis and yield TBEs of $5.09$ and $2.77$ eV for the lowest singlet and triplet transitions. There has been countless experimental and theoretical investigations for this  stable carbene, 
but the most  accurate previous estimates of the vertical transition energies are likely the $5.12$ and $2.83$ eV values, obtained at the MRCI+Q/{\AVTZ} level of theort. \cite{Sun16b}

\emph{Difluorodiazirine.} This cyclopropene analogue is the largest derivative considered herein. There is a remarkable agreement between CC3 and CCSDT values, and the $\Td$ value is very large for each transition, so that we consider the CC values
to obtain our TBEs. For the $^1B_1$ and $^3B_1$ transitions, FCI/{\Pop} calculations deliver transition energies of $3.81\pm0.01$ and $3.09\pm0.01$ eV, perfectly consistent with the present CC values. Our TBEs are likely the most accurate to date for vertical 
transitions.  At the GVVPT2/cc-pVTZ level, the transition energies reported in Ref.~\citenum{Pan04} are $2.25$ eV ($^3B_1$), $2.95$ eV ($^1B_1$), $4.86$ eV ($^3B_2$), $5.21$ eV ($^3A_2$), $6.63$ eV ($^1A_2$), and $8.23$ eV ($^1B_2$), which follows exactly the 
same state ordering as the present CCSDT values. More recently, QCISD/{\AVTZ} estimates of $2.81$ and $3.99$ eV for the lowest triplet and singlet vertical transitions have been reported, which are respectively 
slightly smaller and larger than the present data. There are also quite a few studies of the 0-0 energies of various states for this derivative, both experimentally \cite{Lom69,Hep74,Sie90} and theoretically. \cite{Pan04,Ter16,Loo19a}  

\emph{Formylfluoride.} For this formal intermediate between carbonylfluoride and formaldehyde, we note that the CCSDTQ/{\Pop} values are bracketed by their CC3 and CCSDT counterparts. The previous best estimates are likely the very 
recent MRCI-F12 results of Pradhan and Brown who reported vertical transition energies of $6.03$ eV and $5.68$ eV for the $^1A''$ and $^3A''$ states, respectively. These energies obtained on the CCSD(T)-F12 ground-state geometries are only ca.~$0.05$ eV 
larger than the present TBEs. Most other previous studies focussed on 0-0 energies of the lowest singlet state, \cite{Gid62,Fis69,Sta95b,Cra97,Fan01,Bok09,Loo18b,Loo19a,Pra19} and it is noteworthy that CC3 reproduces the experimental 0-0 energies 
with high accuracy. \cite{Loo18b,Loo19a} Our TBE for the singlet state ($5.96$ eV) is much larger than the measured 0-0 peak ($4.64$ eV) \cite{Cra97} which is expected for a molecule undergoing an important geometrical relaxation after excitation. \cite{Sta95b} 

\emph{HCCl, HCF, and HSiF.} For these three compounds, the SCI calculations deliver values very close to the CC estimates. For \ce{HCCl}, a MRCI+Q/quintuple-$\zeta$ vertical transition energy, corrected for ground-state ZPVE effects, of $1.68$ eV was recently 
reported. \cite{Sha15c} Given that the ZPVE energy at the MP2/{\AVTZ} level is $0.31$ eV, our TBE is basically equivalent to this recent result. For \ce{HSiF}, the most accurate previous estimate of the excitation energy is likely the CC3/{\AVTZ} $3.07$ eV value, 
\cite{Chr02} which is extremely close to our TBE. For the records, Ehara and coworkers also investigated the 0-0 energies and excited-state geometries of these three systems at the SAC-CI level, \cite{Eha11} and experimental 0-0 energies of 
$1.52$ eV (\ce{HCCl}), \cite{Cha95c} $2.14$ eV (\ce{HCF}), \cite{Kak81,Sch99} and $2.88$ eV (\ce{HSiF}), \cite{Har95} have been measured.

\emph{HCP.} Phosphaethyne is a linear compound for which the CC series and the SCI values do converge rapidly and give equivalent results. Consequently, one can trust the TBEs listed in Table \ref{Table-1}. We nevertheless note that there is a 
significant basis set effect for the $^1\Delta$ excited state that is downshifted by $0.05$ eV from {\AVTZ} to \emph{aug}-cc-pV5Z (see Table S1 in the SI). The two most refined previous theoretical works we are aware of have been performed at the 
MRCI/double-$\zeta$ \cite{Nan00} and CC3/cc-pVQZ \cite{Ing06} levels of theory and respectively focussed on reproducing the experimental vibronic couplings and understanding the \ce{HCP -> HPC} isomerization process. However, 
somehow surprisingly, we could not find recent estimates of the vertical transition energies for phosphaethyne, the previously published data being apparently of CASSCF quality.  \cite{Gol93} There are, of course, experimental characterizations 
of the 0-0 energies for several excited states of this compound. \cite{Her66}

\emph{HPO and HPS.} The lowest excited state of HPO has been studied several times in the last twenty years, \cite{Lun95,Tac02,Lee07b,Eha11,Loo19a} whereas its sulfur analogue has only been considered more recently. \cite{Gri13b,Mok14,Meh18,Loo19a} In both cases, 
refined MRCI calculations of the vibronic spectra have been performed \cite{Lee07b,Eha11,Gri13b,Mok14,Meh18} but few reported vertical transition energies. We are aware of a quite old CASPT2 estimate of $2.25$ eV for \ce{HPO},  \cite{Lun95}
and a recent MRCI vertical transition energy of $1.69$ eV (obtained with a very large basis set) for \ce{HPS}. \cite{Meh18} 

\emph{SiCl$_2$.} In this heavier analogue of dichlorocarbene, there are no strong methodological effects but the SCI convergence is shaky, especially for the triplet and we used a basis set extrapolated CCSDTQ value as TBE for this state.
Advanced calculations of the adiabatic energies  \cite{Cha99} as well as experimental 0-0 energies \cite{Du91,Kar93b} can be found in the literature, the latter being $3.72$ and $2.35$ eV for the lowest singlet and triplet states, respectively. These
values are sightly larger than our vertical estimates. For the vertical singlet excitation, there is also a recent $4.06$ eV  CCSD//CAM-B3LYP estimate, \cite{Ran16b} which slightly overshoots ours, consistent with the expected error
sign of CCSD. \cite{Sch08,Kan17,Loo18a}

\emph{Silylidene.} One notes an excellent agreement between CCSDT, CCSDTQ, and FCI for this derivative. Our TBEs of $2.11$ eV and $3.78$ eV are again exceeding the experimental 0-0 energies of $1.88$ eV \cite{Smi03} 
and $3.63$ eV, \cite{Har97b} as it should. The previous theoretical studies we are aware of have been performed with CISD(+Q), \cite{Hil97,Smi03} and CC3 \cite{Loo18b,Loo19a} methods and mainly discussed the 0-0 energies, for which an 
excellent agreement with experiment was obtained by both approaches.

\subsubsection{Benchmarks}

\begin{figure*}[htp]
  \includegraphics[scale=0.85,viewport=2.8cm 5.5cm 18.3cm 27.5cm,clip]{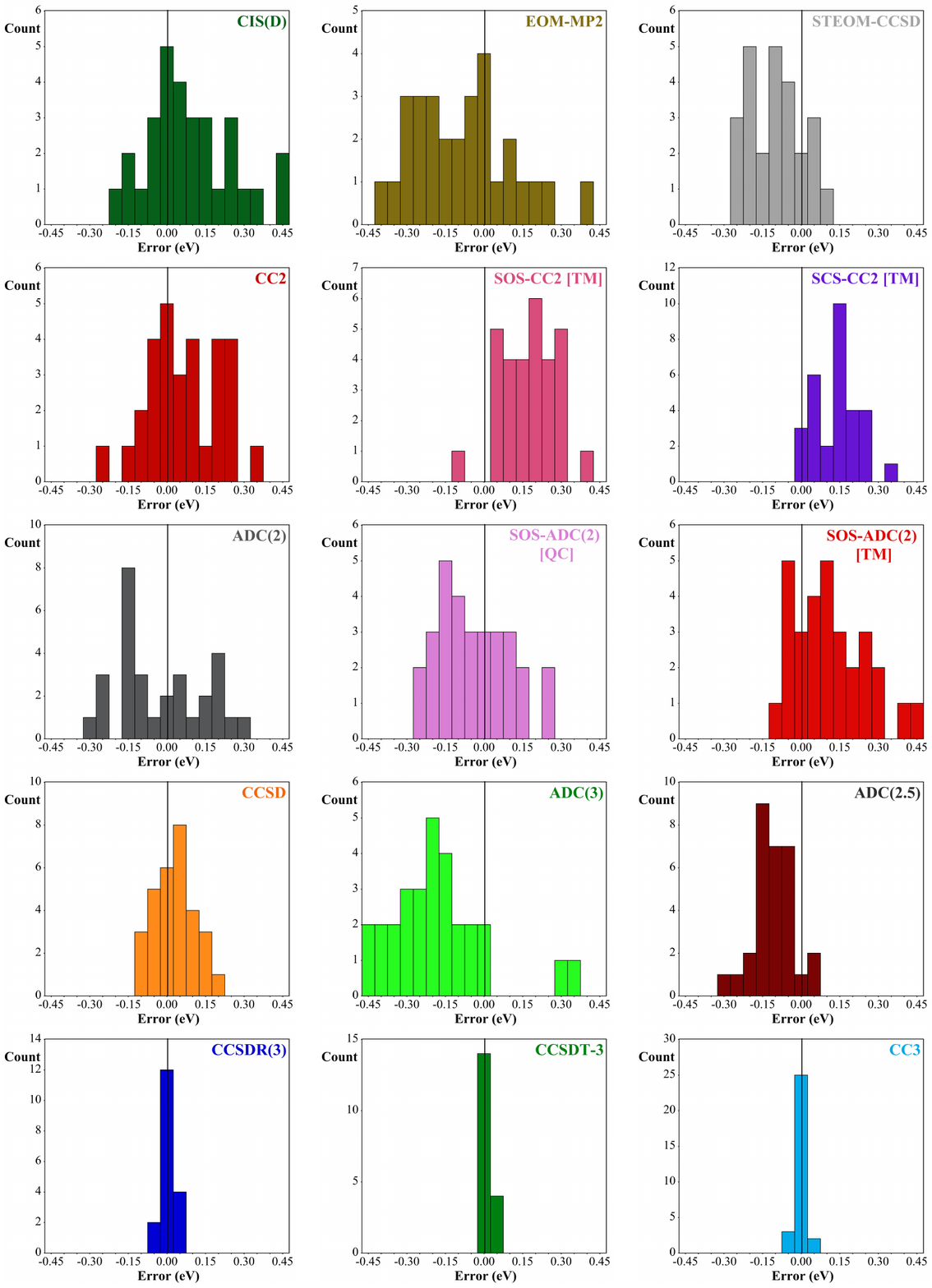}
  \caption{Histograms of the error distribution (in eV) obtained with 15 theoretical methods, choosing the TBE/{\AVTZ} of Table \ref{Table-1} as references.
  TM and QC stand for the TURBOMOLE and Q-CHEM definitions of the scaling factors, respectively.
Note the difference of scaling in the vertical axes.}
  \label{Fig-1}
\end{figure*}

Benchmarks using the TBEs obtained in the previous Section can be naturally done. As we consider closed-shell compounds, there is a large number of methods that one can evaluate. Here, we have chosen 15 popular wave function 
methods for excited states (see \textit{Computational Details} and Table S2 in the SI for the raw data). The statistical result can be found in Figure \ref{Fig-1} and Table \ref{Table-2}.

\renewcommand*{\arraystretch}{1.0}
\begin{table}[htp]
\scriptsize
\caption{Statistical values obtained by comparing the results of various methods to the TBE/{\AVTZ} values listed in Table \ref{Table-1}. We report the 
mean signed error (MSE), mean absolute error (MAE), root-mean square error (RMSE), and standard deviation of the errors (SDE). All 
quantities are given in eV and have been obtained with the {\AVTZ} basis set. TM and QC stand for the TURBOMOLE and Q-CHEM
definitions of the scaling factors, respectively. ADC(2.5) is the simple average of the ADC(2) and ADC(3) transition energies, as defined in Ref.~\citenum{Loo20b}.
``Count'' refers to the number of transitions computed for each method.}
\label{Table-2}
\begin{tabular}{lccccc}
\hline
Method 		& Count 	& MSE 	&MAE 	&RMSE 	&SDE 		\\
\hline
CIS(D)			&30&0.09		&0.14	&0.19	&0.16 		\\
EOM-MP2			&30&-0.06	&0.17	&0.22	&0.21 		\\
STEOM-CCSD		&25&-0.10	&0.12	&0.14	&0.10 		\\
CC2				&30&0.07		&0.12	&0.15	&0.14 		\\
SOS-CC2 [TM]		&30&0.17		&0.18	&0.20	&0.10 		\\
SCS-CC2 [TM]		&30&0.14		&0.14	&0.16	&0.09 		\\
ADC(2)			&30&-0.02	&0.15	&0.16	&0.17 		\\
SOS-ADC(2) [TM]	&30&0.11		&0.13	&0.17	&0.14 		\\
SOS-ADC(2) [QC]	&30&-0.04	&0.12	&0.14	&0.14 		\\
CCSD			&30&0.03		&0.07	&0.08	&0.08 		\\
ADC(3)			&30&-0.19	&0.24	&0.27	&0.19 		\\
ADC(2.5)			&30&-0.11		&0.11	&0.13	&0.07 		\\
CCSDR(3)		&19&0.01		&0.02	&0.02	&0.02 		\\
CCSDT-3			&19&0.01		&0.02	&0.02	&0.02 		\\
CC3				&30&0.00		&0.01	&0.02	&0.02 		\\
\hline
 \end{tabular}
 \end{table}

Most of the conclusions that can be extracted from these benchmarks are consistent with recent analyses made in the field, \cite{Kan14,Taj16,Kan17,Loo18a,Loo19a,Loo20a,Taj20a,Loo20b} and we will therefore only briefly comment on the most 
significant outcomes. First, one notes that CC3, which is an expensive approach, is superbly accurate and consistent with a trifling MSE and a tiny SDE, whereas both CCSDT-3  and CCSDR(3), for which only singlet excited states can be 
evaluated with the current implementations, are also extremely satisfying with average errors well below the chemical accuracy threshold. This is unsurprisingly inline with the trends obtained for more ``standard'' organic compounds: CC 
methods including (at least partially) contributions from the triples are trustworthy for the description of single excitations. \cite{Hat05c,Sau09,Wat13,Kan17,Loo18a,Loo19a,Sue19} Going down in the CC hierarchy, we find that CCSD slightly 
overestimates the transition energies, but nevertheless provides very consistent estimates (SDE of $0.08$ eV), whereas CC2 is clearly less satisfying in terms of consistency (SDE of $0.14$ eV).  Comparing with previous benchmarks, 
\cite{Sch08,Car10,Wat13,Kan14,Jac17b,Kan17,Dut18,Jac18a,Loo18a,Loo20b} we can foresee that the CCSD overestimation will likely grow in larger compounds, whereas the CC2 accuracy should remain less affected by the system size. The SOS and SCS 
variants of CC2 deliver larger MAE, with a clear overestimation (see Figure \ref{Fig-1}), but a smaller error dispersion than the standard CC2 method. The accuracy deterioration and the improved consistency of the spin-scaled CC2 
versions (w.r.t. standard CC2) is known, \cite{Goe10a,Jac15b,Taj20a} though some works reported that SOS-CC2 and SCS-CC2 can also improve the accuracy.\cite{Win13} STEOM-CCSD delivers results of roughly CC2 quality for the present set, whereas patterns 
more alike the ones of CCSD have been previously obtained. \cite{Loo18a,Dut18,Loo20a} In the present case, both CIS(D) and EOM-MP2 [also denoted CCSD(2)], which are the two computationally lightest approaches, are also the ones 
yielding the largest dispersions alongside quite significant MAEs.  For EOM-MP2, similar outcomes were observed for valence excited states by Tajti and Szalay, \cite{Taj16} whereas the relatively poor performance of CIS(D) is well 
documented. \cite{Goe10a,Jac15b,Loo18a,Loo20a} In the ADC series, we note that ADC(2) yields results only slightly less accurate than CC2 for a smaller computational cost, which is consistent with the conclusions of Dreuw's group, \cite{Har14} 
whereas the SOS variant developed by the same group \cite{Kra13} has a slight edge over its TURBOMOLE variant. ADC(3) provides rather poor excitation energies, a trend we recently evidenced in other molecular sets. 
\cite{Loo18a,Sue19,Loo20b} Finally, the very recently introduced ADC(2.5) scheme, which corresponds to the simple average of the ADC(2) and ADC(3) excitation energies, \cite{Loo20b} provides significantly more consistent estimates 
than both ADC(2) or ADC(3), with a SDE of $0.07$ eV only compared to ca.~$0.18$ eV for the ``parent'' methods. ADC(2.5) can then be seen as a cost effective approach to improve upon ADC(3), at least for small compounds.

\subsection{Radical set}

\subsubsection{Reference values and comparison to literature}

Let us now turn to radicals. As nicely summarized by Crawford fifteen years ago, \cite{Smi05b} electronic transitions in open-shell systems are more challenging, not only due to the more limited number of methods and codes available for treating them 
(as compared to closed-shell molecules), but also because: (i) strong spin contamination can take place with ``low''-level methods; (ii) large contributions from doubly-excited configurations are quite common; and (iii) basis set effects can be 
very large, meaning that reaching the CBS limit can be laborious. At the CCSD level for instance, significant differences between U and RO transition energies can sometimes be observed. \cite{Smi05b} This is why our results, listed in Table \ref{Table-3}, 
use as computationally-lightest approach (U)CCSDT, so that the wave function is robust enough in order to mitigate the two former issues for most of the considered transitions.  As can be seen in the {\Pop} and {\AVDZ} columns of Table \ref{Table-3}, 
one generally finds an excellent agreement between the various CC estimates and their FCI counterparts, UCCSDT being already extremely accurate except in specific cases (such as the $^2\Sigma^+$ excited state of \ce{CO+}). This overall consistency 
indicates yet again that one can trust the present TBEs. We underline that, except for diatomics, UCCSDT calculations performed with diffuse basis sets on open-shell molecules are quite rare in the literature 
(see below), and the same obviously holds for higher-order CC. As for the exotic set, we do not intend here to provide an exhaustive list of previous works, but rather to pinpoint a few interesting comparisons with earlier accurate estimates.

\begin{table*}[htp]
\scriptsize
\caption{Excitation energies (in eV) of the radical set obtained within the FC approximation. For each state, we report, on the left hand side, the TBE/{\AVTZ} excitation energy obtained directly at the FCI level (except otherwise stated). The TBE/CBS excitation energy is obtained with the largest affordable basis set (see footnotes). On the right hand side, one finds the transition energies computed at various levels of theory.  T, TQ, and TQP stand for UCCSDT, UCCSDTQ, and UCCSDTQP, respectively.}
\label{Table-3}
\vspace{-0.3 cm}
\begin{tabular}{ll|ll|p{.36cm}p{.36cm}p{.36cm}p{1.0cm}|p{.36cm}p{.36cm}p{.36cm}p{1.0cm}|p{.36cm}p{.36cm}p{1.0cm}|p{.36cm}p{.36cm}p{1.0cm}}
\hline 
			&			& AVTZ	& CBS	&	 \multicolumn{4}{c}{\Pop} 					& \multicolumn{4}{c}{\AVDZ}  				& \multicolumn{3}{c}{\AVTZ}		& \multicolumn{3}{c}{\AVQZ} \\
			&			& TBE	& TBE	&	T	&	 TQ		& TQP	& FCI		& T	&	 TQ		& TQP	& FCI		& T	&	 TQ		& FCI		&  T	&	 TQ		& FCI	 \\
\hline
Allyl			&$^2B_1$		&3.39$^a$&		&	3.46&	3.44&		&	3.42$\pm$0.02	&3.46&		&		&	3.44$\pm$0.04&3.43		&				&		&		&						\\
			&$^2A_1$		&4.99$^a$&		&	5.16&	5.14&		&	5.18$\pm$0.01	&4.88&		&		&	4.91$\pm$0.04&4.97		&				&		&		&						\\
\ce{BeF}		&$^2\Pi$		&4.14	&4.13$^b$&	4.29&	4.28&	4.28&	4.28$\pm$0.00	&4.21&	4.20&	4.20	&	4.20$\pm$0.09	&4.15&	4.15&	4.14$\pm$0.01	&	4.14&		&	4.13$\pm$0.01		\\
			&$^2\Sigma^+$	&6.21	&		&	6.32&	6.31&	6.31&	6.32$\pm$0.00	&6.30&	6.29&	6.29	&	6.29$\pm$0.00	&6.23&	6.22&	6.21$\pm$0.02	&		&		&					\\
\ce{BeH}		&$^2\Pi$		&2.49	&2.48$^b$&	2.53&	2.53&	2.53&	2.53$\pm$0.00	&2.52&	2.52&	2.52&	2.52$\pm$0.00	&2.49&	2.49&	2.49$\pm$0.00 &	2.48&	2.48&	2.48$\pm$0.00		\\
			&$^2\Pi$		&6.46	&6.46$^b$&	6.42&	6.42&	6.42&	6.42$\pm$0.00	&6.43&	6.43&	6.43&	6.43$\pm$0.00	&6.45&	6.46&	6.46$\pm$0.00 &	6.46&	6.46&	6.46$\pm$0.00		\\
\ce{BH2}		&$^2B_1$		&1.18	&1.18$^c$&	1.19&	1.19&	1.19&	1.19$\pm$0.00	&1.21&	1.21&	1.21&	1.21$\pm$0.00	&1.18&	1.18&	1.18$\pm$0.01	&	1.18&	1.18&	1.18$\pm$0.00		\\
\ce{CH}		&$^2\Delta$	&2.91	&2.90$^c$&	3.07&	3.05&	3.05&	3.05$\pm$0.00	&3.01&	2.99&	2.99&	2.99$\pm$0.00	&2.94&	2.91&	2.91$\pm$0.00	&	2.93&	2.90&	2.90$\pm$0.00		\\
			&$^2\Sigma^-$	&3.29	&3.28$^c$&	3.36&	3.35&	3.35&	3.35$\pm$0.00	&3.34&	3.32&	3.32&	3.32$\pm$0.00	&3.31&	3.29&	3.29$\pm$0.00	&	3.30&	3.29&	3.28$\pm$0.01		\\
			&$^2\Sigma^+$	&3.98	&3.96$^c$&	4.12&	4.10&	4.10&	4.09$\pm$0.00	&4.07&	4.04&	4.04&	4.03$\pm$0.00	&4.03&	3.98&	3.98$\pm$0.00	&	4.01&	3.97&	3.96$\pm$0.01		\\
\ce{CH3}		&$^2A_1'$	&5.85	&5.88$^c$&	6.00&	6.00&	6.00&	6.00$\pm$0.00	&5.78&	5.79&	5.79&	5.79$\pm$0.00	&5.86&	5.86&	5.85$\pm$0.01	&	5.88&		&	5.88$\pm$0.00		\\
			&$^2E'$		&6.96	&6.96$^c$&	7.28&	7.28&	7.28&	7.28$\pm$0.00	&7.01&	7.02&	7.02&	7.01$\pm$0.00	&6.97&	6.97&	6.96$\pm$0.01	&	6.96&		&	6.96$\pm$0.00		\\
			&$^2E'$		&7.18	&7.17$^d$&	7.43&	7.43&	7.43&	7.43$\pm$0.00	&7.17&	7.18&	7.18&	7.18$\pm$0.00	&7.19&	7.19&	7.18$\pm$0.02	&	7.19&		&					\\
			&$^2A_2''$	&7.65	&7.48$^d$&	7.81&	7.81&	7.81&	7.81$\pm$0.00	&7.76&	7.76&	7.76&	7.76$\pm$0.00	&7.65&	7.66&	7.65$\pm$0.01	&	7.57&		&					\\
\ce{CN}		&$^2\Pi$		&1.34	&1.33$^b$&	1.44&	1.41&	1.41&	1.40$\pm$0.00	&1.42&	1.39&	1.38	&	1.38$\pm$0.00	&1.38&	1.35	&	1.34$\pm$0.01	&	1.38&		&	1.33$\pm$0.01		\\
			&$^2\Sigma^+$	&3.22	&3.21$^b$&	3.24&	3.23&	3.23&	3.23$\pm$0.01	&3.25&	3.23&	3.23	&	3.23$\pm$0.00	&3.25&	3.22	&	3.22$\pm$0.00	&	3.25&		&	3.21$\pm$0.00		\\
\ce{CNO}		&$^2\Sigma^+$	&1.61	&1.61$^e$&	1.66&	1.59&		&	1.57$\pm$0.00	&1.66&	1.59	&		&	1.58$\pm$0.00	&1.71&		&	1.61$\pm$0.01	&	1.71	&		&					\\
			&$^2\Pi$		&5.49$^a$&5.50$^f$&	5.62&	5.56&		&	5.54$\pm$0.02	&5.59& 	5.53&		&	5.49$\pm$0.05	&5.57&		&				&	5.58	&		&					\\
\ce{CON}		&$^2\Pi$$^g$	&3.53$^h$&		&	3.54	&	3.50	&		&	3.53$\pm$0.00	&3.55&		&		&	3.54$\pm$0.01	&3.54&		&		&		&					\\
		&$^2\Sigma^+$	$^g$	&3.87$^i$	&		&	4.04	&	3.79&		&				&4.05&		&		&				&4.12&		&		&		&					\\
\ce{CO+}		&$^2\Pi$		&3.28	&3.26$^b$&	3.30&	3.33&	3.33&	3.33$\pm$0.00	&3.30&	3.33&	3.33	&	3.33$\pm$0.00	&3.26&	3.28	&	3.28$\pm$0.00	&	3.27&		&	3.26$\pm$0.00		\\
			&$^2\Sigma^+$	&5.81	&5.80$^b$&	5.69&	5.79&	5.82&	5.82$\pm$0.00	&5.78&	5.87&	5.89	&	5.90$\pm$0.00	&5.70&	5.78	&	5.81$\pm$0.00	&	5.72&		&	5.80$\pm$0.00		\\
\ce{F2BO}		&$^2B_1$		&0.73$^h$&		&	0.73&		&		&	0.72$\pm$0.00	&0.72&		&		&	0.74$\pm$0.02	&0.71&		&				&		&		&					\\
			&$^2A_1$		&2.80$^h$&		&	2.85&		&		&	2.87$\pm$0.00	&2.86&		&		&	2.88$\pm$0.00	&2.78&		&				&		&		&					\\
\ce{F2BS}		&$^2B_1$		&0.51$^h$&		&	0.49&		&		&	0.48$\pm$0.00	&0.50&		&		&	0.53$\pm$0.00	&0.48&		&				&		&		&					\\
			&$^2A_1$		&2.99$^h$&		&	3.07&		&		&	3.06$\pm$0.03	&2.96&		&		&	3.02$\pm$0.01	&2.93&		&				&		&		&					\\
\ce{H2BO}	&$^2B_1$		&2.15	&2.14$^e$&	2.27&	2.28&	2.28&	2.28$\pm$0.00	&2.23&	2.23&		& 	2.23$\pm$0.00	&2.17&		&	2.15$\pm$0.01	&	2.16&		&				\\
			&$^2A_1$		&3.49	&3.49$^e$&	3.62&	3.62&	3.62&	3.62$\pm$0.00	&3.61&	3.61&		&	3.60$\pm$0.01	&3.51&		&	3.49$\pm$0.01	&	3.51&		&				\\
\ce{HCO}		&$^2A''$		&2.09	&2.09$^e$&	2.18&	2.17&	2.18&	2.17$\pm$0.01	&2.12&	2.12&		&	2.13$\pm$0.01	&2.10&		&	2.09$\pm$0.01	&	2.10&		&				\\ 
			&$^2A'$		&5.45$^h$&5.49$^j$&	5.45&	5.47&	5.47&	5.47$\pm$0.00	&5.32&	5.33&		&	5.33$\pm$0.01	&5.44&		&	5.42$\pm$0.07	&	5.48&		&				\\
\ce{HOC}		&$^2A''$		&0.92	&0.91$^e$&	0.99&	0.99&	0.99&	0.99$\pm$0.00	&0.96&	0.96&		&	0.96$\pm$0.00	&0.93&		&	0.92$\pm$0.00	&	0.92&		&				\\
\ce{H2PO}	&$^2A''$		&2.80	&2.83$^e$&	2.85&	2.86&	2.86	&	2.88$\pm$0.01	&2.80&	2.82	&		&	2.82$\pm$0.02	&2.81&		&	2.80$\pm$0.02 &	2.84&		&				\\
			&$^2A'$		&4.21$^h$&4.22$^j$&	4.30&	4.30&	4.30&	4.31$\pm$0.00	&4.28&	4.28&		&	4.28$\pm$0.02	&4.21&		&	4.19$\pm$0.04	&	4.22&		&				\\
\ce{H2PS}		&$^2A''$		&1.16	&1.18$^e$ &	1.10&	1.10&	1.10&	1.11$\pm$0.00	&1.16&	1.16	&		&	1.17$\pm$0.00	&1.15&		&	1.16$\pm$0.01	&	1.17&		&				\\
			&$^2A'$		&2.72	&2.71$^e$ &	2.88&	2.87&	2.87&	2.87$\pm$0.00	&2.81&	2.80	&		&	2.80$\pm$0.00	&2.75&		&	2.72$\pm$0.02	&	2.74&		&				\\
\ce{NCO}		&$^2\Sigma^+$	&2.89$^h$&2.89$^j$&	2.87&	2.87&		&	2.88$\pm$0.00	&2.87&	2.86	&		&	2.89$\pm$0.02	&2.87&		&	2.83$\pm$0.05	&	2.87	&		&				\\
			&$^2\Pi$		&4.73$^h$&4.74$^j$&	4.80&	4.76&		&	4.76$\pm$0.00	&4.80&	4.76	&		&	4.76$\pm$0.01	&4.77&		&	4.70$\pm$0.04 &	4.78	&		&				\\
\ce{NH2}		&$^2A_1$		&2.12	&2.11$^c$	&	2.19&	2.18&	2.18&	2.18$\pm$0.00	&2.15&	2.15&	2.15&	2.14$\pm$0.00	&2.12&	2.12&	2.12$\pm$0.00	&	2.11&	2.11	&	2.11$\pm$0.00\\
Nitromethyl	&$^2B_2$		&2.05$^k$&		&	2.10&		&		&				&2.04&		&		&				&2.05&		&				&		&		&					\\
			&$^2A_2$		&2.38$^k$&		&	2.40&		&		&	2.39$\pm$0.01	&2.39&		&		&				&2.38&		&				&		&		&					\\
			&$^2A_1$		&2.56$^k$&		&	2.64&		&		&				&2.58&		&		&				&2.56&		&				&		&		&					\\
			&$^2B_1$		&5.35$^k$&		&	5.48&		&		&				&5.39&		&		&				&5.35&		&				&		&		&					\\
\ce{NO}		&$^2\Sigma^+$&6.13	&6.12$^e$&	6.12&	6.11&	6.11&	6.11$\pm$0.00	&6.03&	6.02&	6.02&	6.03$\pm$0.01	&6.13&	6.12&	6.13$\pm$0.02	&	6.12	&		&	\\
			&$^2\Sigma^+$	&7.29$^l$&7.21$^m$&	7.59&	7.59&	7.59&				&7.34&	7.34&	7.34&				&7.29&	7.29&				&	7.21	&		&	\\
\ce{OH}		&$^2\Sigma^+$	&4.10	&4.09$^c$&	4.28&	4.28&	4.28&	4.28$\pm$0.00	&4.16&	4.16&	4.16&	4.16$\pm$0.00	&4.12&	4.12&	4.10$\pm$0.01	&	4.11&	4.10&	4.10$\pm$0.00\\
			&$^2\Sigma^-$	&8.02	&8.11$^c$&	8.83&	8.83&	8.83&	8.83$\pm$0.00	&7.88&	7.88&	7.88&	7.88$\pm$0.00	&8.04&	8.02&	8.02$\pm$0.00	&	8.10&	8.09&	8.09$\pm$0.00\\
\ce{PH2}		&$^2A_1$		&2.77	&2.76$^c$&	2.90&	2.90&	2.90&	2.90$\pm$0.00	&2.79&	2.79&	2.79&	2.79$\pm$0.00	&2.77&	2.77&	2.77$\pm$0.00	&	2.76&	2.76	&	2.76$\pm$0.00\\
Vinyl			&$^2A''$		&3.26	&		&	3.45&	3.43&	3.43&	3.43$\pm$0.00	&3.36&	3.34&		&	3.35$\pm$0.00	&3.31&		&	3.26$\pm$0.02 &		&		&				\\
			&$^2A''$		&4.69	&		&	4.98&	4.96&	4.96&	4.96$\pm$0.00	&4.80&		&		&	4.78$\pm$0.01	&4.73&		&	4.69$\pm$0.02	& \\
			&$^2A'$$^g$	&5.60	&		&	5.83	&	5.75	&	5.75&	5.74$\pm$0.01	&5.75&	5.67&		&	5.68$\pm$0.00	&5.74 &		&	5.60$\pm$0.01	& \\
			&$^2A'$		&6.20$^a$&		&	6.50	&	6.48	&		&	6.49$\pm$0.01	&6.15&	6.14&		&				& 6.21&		&				&	\\
\hline			
\end{tabular}
\vspace{-0.3 cm}
\begin{flushleft}
$^a${FCI/{\Pop} value corrected by the difference between CCSDT/{\AVTZ} and CCSDT/{\Pop};}
$^b${FCI/{\AVQZ} value;}
$^c${FCI/{\AVQZ} value corrected by the difference between CCSDT/{\AVFZ} and CCSDT/{\AVQZ};}
$^d${FCI/{\AVTZ} value corrected by the difference between CCSDT/{\AVFZ} and CCSDT/{\AVTZ};}
$^e${FCI/{\AVTZ} value corrected by the difference between CCSDT/{\AVQZ} and CCSDT/{\AVTZ};}
$^f${FCI/{\Pop} value corrected by the difference between CCSDT/{\AVQZ} and CCSDT/{\Pop};}
$^g${For these challenging states, ROCC rather than UCC is used.}
$^h${FCI/{\AVDZ} value corrected by the difference between CCSDT/{\AVTZ} and CCSDT/{\AVDZ};}
$^i${CCSDTQ/{\Pop} value corrected by the difference between CCSDT/{\AVTZ} and CCSDT/{\Pop};}
$^j${FCI/{\AVDZ} value corrected by the difference between CCSDT/{\AVQZ} and CCSDT/{\AVDZ};}
$^k${CCSDT/{\AVTZ} value;}
$^l${CCSDTQ/{\AVTZ} value;}
$^m${CCSDTQ/{\AVTZ} value corrected by the difference between CCSDT/{\AVQZ} and  CCSDT/{\AVTZ}.}
\end{flushleft}
\end{table*}

\emph{Allyl.} For the lowest valence ($B_1$) and Rydberg ($A_1$) transitions of the allyl radical, the previous TBEs are likely the ROCC3 $3.44$ and $4.94$ eV vertical transition energies obtained by the Crawford group with the {\AVTZ} basis further
augmented with molecule-centered functions (mcf). \cite{Mac10} For the lowest state, a very similar value of $3.43$ eV was obtained at the ROCC3 level without mcf. \cite{Loo19a} The present work is the first to report 
CCSDT and CCSDTQ results. They clearly show that these previous ROCC3 estimates are very accurate. In addition, our TBEs of $3.39$ and $4.99$ eV are reasonably consistent  with earlier CASPT2 ($3.32$ and $5.11$ eV) 
\cite{Aqu03b} and MRCI ($3.32$ and $4.68$ eV) \cite{Gas10} data. The experimental 0-0 energies have been reported to be $3.07$ eV, \cite{Cas06c} and $4.97$ eV \cite{Gas09,Gas10} for the $^2B_1$ and $^2A_1$ states, respectively. 
The fact that the experimental $T_0$ value is very close to the computed vertical transition energy of the second state is rather surprising, but remains unchanged with the present work.

\emph{BeF.} In this compound, CCSDT delivers transition energies in very good agreement with FCI (and higher CC levels), but one notices a non-negligible basis set effect for the second transition of Rydberg character. This transition
becomes significantly mixed in very large basis sets, making a clear attribution difficult.  For this derivative (and other diatomics), experimental vertical transition energies can be calculated by analyzing the experimental spectroscopic 
constants. \cite{Mau95} Our TBE/{\AVTZ} values of $4.14$ and $6.21$ eV are obviously close to these measured values of $4.14$ and $6.16$ eV. \cite{Mau95} For the lowest state, a previous MRCI value of $4.23$ eV can be 
found in the literature. \cite{Orn92} There is also a recent evaluation of the adiabatic energies for numerous excited states at the MRCI+Q level. \cite{Elk17}

\emph{BeH.} The convergences with respect to both the CC excitation order and the basis set size is extremely fast for this five-electron system. A previous study reports FCI values for many excited states \cite{Pit08} and, in particular, excitation energies of $2.53$ and $6.30$ eV 
for the two $^2\Pi$ states considered herein. The experimental vertical transition energies are $2.48$ and $6.32$ eV. \cite{Mau95} Our larger value associated with the second transition is likely a consequence of the UCCSD(T) 
geometry, which delivers a slightly shorter bond length ($1.321$ \emph{vs} $1.327$ \AA\ experimentally).
 
\emph{BH$_2$, NH$_2$, and PH$_2$.}  In these three related compounds, convergence with respect to the CC excitation order and basis set size is also very fast, so that accurate estimates can be easily produced for the lowest-lying transition: 
With the {\AVTZ} basis set, near-CBS excitation energies of $1.18$, $2.12$ and $2.77$ eV for the boron, nitrogen, and phosphorus derivative are respectively obtained. For \ce{BH2}, a previous MRCI estimate of $1.10$ eV is available in the literature. \cite{Per95} 
We note that, for \ce{BH2}, the geometry relaxation of the bent ground state structure would lead to a linear geometry in its lowest excited state, \cite{Sun15d}  a phenomenon that was extensively studied both experimentally and 
theoretically (see Ref.~\citenum{Sun15d} and references therein).  For \ce{NH2}, a vertical estimate of $2.18$ eV was reported by Szalay and Gauss using a CCSD approach  including ``pseudo'' triple excitations, \cite{Sza00} and 
high-order CC calculations have been latter performed by Kallay and Gauss to investigate the structures and energetics of the ground and excited states. \cite{Kal03,Kal04} For \ce{PH2}, the most detailed \emph{ab initio} studies 
that are available in the literature focus exclusively on the 0-0 energies and rovibronic spectra, \cite{Woo01,Yur06,Jak06} except for a recent report listing a ROCC3 vertical transition energy of $2.75$ eV, \cite{Loo19a} obviously close to present TBE.

\emph{CH.} For the three considered transitions, the CCSDT values are slightly too large, whereas the basis set effects are rather usual, with nearly converged results for the {\AVTZ} basis set. Although we consider a theoretical geometry, our basis set corrected 
TBEs of $2.90$, $3.28$, and $3.96$ eV for the $^2\Delta$, $^2\Sigma^-$, and $^2\Sigma^+$ states are all extremely close to the vertical experimental values of $2.88$, $3.26$ and $3.94$ eV. \cite{Mau95,Sli05} 
There are many previous works on the \ce{CH} radical and it is interesting to mention that the ROCCSD values are $3.21$, $4.25$, and $5.22$ eV for the same three states, \cite{Sza00}  whereas the 
corresponding ROCC3 results are $3.16$, $3.58$, and $4.47$ eV; \cite{Smi05b} the ROCC(2,3) excitation energies are $2.97$, $3.33$, and $4.06$ eV. \cite{Sli05} This clearly illustrates the challenge of reaching accurate values for
the second and third transitions with ``low-order'' methods.  For \ce{CH},  high-order CC calculations of the adiabatic energies and other properties are also available in the literature. \cite{Hir04,Fan07}

\emph{CH$_3$.}  For the methyl radical, the convergence of the CC excitation energies and the near-perfect agreement between CC and FCI is worth noting.
Nonetheless,  large basis set effects are present for these transition energies, especially for the high-lying $^2A_2''$
state for which the {\AVTZ} excitation energy is still far from being converged basis set wise.  Our TBEs, including corrections up to quintuple-$\zeta$ are: $5.88$, $6.96$, $7.17$, and $7.48$ eV for the four lowest transitions. These values can be compared
to the previous MRCI estimates \cite{Meb97b,Zan16} of $5.86$ ($5.91$), $6.95$ ($7.03$), $7.13$ (--) and $7.37$ (7.66) eV reported in Ref.~\citenum{Meb97b} (\citenum{Zan16}). The experimental $T_0$ value is $5.73$ eV for the
$^2A_1'$ state, \cite{Her66,Set03b} whereas the experimental $T_e$ value is $7.43$ eV for the $^2A_2''$ state,  \cite{Hud83,Fu05} both slightly below our FCI vertical estimates.

\emph{CN.} Both methodological and basis set effects are firmly under control for the cyano radical, so that our FCI/{\AVTZ} results of $1.34$ and $3.22$ eV for the lowest excited states are likely very accurate for the considered geometry. 
These values are indeed close to the experimental energies of $1.32$ and $3.22$ eV. \cite{Mau95} One can find careful MRCI studies, \cite{Yaz05,Shi11} as well as an extensive benchmark \cite{Bao17}
for the adiabatic energies of the \ce{CN} radical.

\emph{CNO, CON, and NCO.} Inspired by a previous investigation, \cite{Yaz05} we have evaluated the two lowest doublet transitions in these three linear isomers. For \ce{CNO} --- the second most stable isomer --- one notes non-negligible 
drops of the transition energies going from CCSDT to CCSDTQ, the latter theory providing data in perfect match with the FCI results. Our TBEs of $1.61$ eV ($^2\Sigma^+$) and $5.50$ eV ($^2\Pi$), do compare very favorably with the 
corresponding MRCI+Q results of $1.66$ and $5.50$ eV, respectively. \cite{Yaz05} For the former transition, there is also a ROCC3 vertical transition energy of $1.71$ eV \cite{Loo19a} and a detailed rovibronic 
investigation \cite{Leo08} available in the literature. The data are much scarcer for \ce{CON}, and the only previous work we are aware of reports potential energy surfaces  without listing explicitly the transition energies. \cite{Yaz05}
For \ce{CON}, we have performed multi-reference calculations to identify the lowest states (see Table S4 in the SI). The NEVPT2 calculations locate the $^2\Pi$ and $^2\Sigma^+$ transitions at $3.52$ and $3.81$ eV, respectively,
similar values being obtained with both CASPT2 and MRCI. As can be seen in Table \ref{Table-3} the FCI-based estimate of $3.53$ eV for the former transition is extremely consistent. For the latter transition, the difference
between CCSDT and CCSDTQ energies is as large as -0.25 eV, suggesting that further corrections would be required. Nevertheless, our CC-derived TBE of $3.87$ eV is rather consistent with the NEVPT2 and MRCI values.
For \ce{NCO}, the most stable of the three isomers, the basis set effects are trifling, but CCSDTQ is again mandatory in order to obtain a very accurate transition energy for the  $^2\Pi$ state. This compound was studied previously at the 
MRCI+Q level, a method which delivers respective vertical transition energies of $2.89$ and $4.68$ eV for the $^2\Sigma^+$ and $^2\Pi$ states, \cite{Yaz05} whereas the ROCC3/{\AVTZ} transition energy of 
the lowest excited state is $2.83$ eV.  \cite{Loo19a} The measured experimental 0-0 energies are $2.82$, \cite{Wu92} and $3.94$ eV. \cite{Dix60} All these data are quite consistent with our new values of $2.89$ and $4.74$ eV.

\emph{CO$^+$.}   Our FCI/{\AVQZ} values for the $^2\Pi$ and $^2\Sigma^+$ transitions, $3.26$ and $5.80$ eV, are clearly matching the experimental values of $3.26$ and $5.81$ eV. \cite{Mau95}  While 
basis set effects are rather standard for this radical cation, it is noteworthy that the CC expansion converges slowly for the Rydberg $^2\Sigma^+$ transition: one needs CCSDTQP to be within $0.01$ eV of the 
FCI result! Nonetheless, previous ROCC3 ($3.29$ and $5.73$ eV) \cite{Smi05b} and ROCC(2,3) data ($3.35$ and $5.81$ eV), \cite{Sli05} also fall within $\pm0.10$ eV of the present TBEs.

\emph{F$_2$BO and F$_2$BS.}  These two radicals present a very low-lying $\pi-n$ transition, that is described very similarly by all basis sets used in Table \ref{Table-3}. For these transitions our
TBEs are $0.73$ (\ce{F2BO}) and $0.51$ (\ce{F2BS}) eV, whereas, for the second transition of $\sigma-n$ nature, our TBEs are $2.80$ (F$_2$BO) and  $2.99$ (\ce{F2BS}) eV.
For these two compounds, the most advanced previous calculations are likely the ROCC3/{\AVTZ} values of $0.71$ and $2.78$ eV (F$_2$BO), and $0.47$ and $2.93$ eV (F$_2$BS)
obtained by some of us in a recent study. \cite{Loo19a} For the former radical, these values are also very close to earlier CASPT2 ($0.70$ and $2.93$ eV) \cite{Bar08b} and SAC-CI ($0.73$ and $2.89$ eV) \cite{Li18b} estimates.
The $T_0$ energies of these two states were both measured recently as well: $0.65$ and $2.78$ eV for the oxygen derivative, \cite{Gri14} and $0.44$ and $2.87$ eV for the sulfur radical. \cite{Jin15} These two
works and an earlier study by the same group, \cite{Clo14} also provide advanced theoretical studies of both the 0-0 transitions and vibronic couplings.

\emph{H$_2$BO.} This lighter analogue of \ce{F2BO} remains to be detected experimentally, but its excited states have been studied twice with \emph{ab initio} theoretical methods, \cite{Clo14,Li18b} the most recent SAC-CI estimates for
the lowest-lying transitions being $2.08$ and $3.49$ eV. \cite{Li18b} These SAC-CI excitation energies are within $0.10$ eV of our FCI-based TBEs.

\emph{HCO and HOC (formyl and isoformyl).} For the formyl radical, our TBEs are $2.09$ and $5.49$ eV. Kus and Bartlett reported CCSDT/6-311++G(d,p) transition energies of $2.17$ and $5.29$ eV (likely the best vertical estimates available previously), \cite{Kus08} obviously close to ours
for the former valence transition. We are also aware of earlier CASPT2 estimates of $2.07$ and $5.45$ eV for these two states, \cite{Ser98} that happen to be within
$\pm0.04$ eV of our TBEs. There are detailed studies of the potential energy surfaces for the ground and lowest excited states of \ce{HCO}. \cite{Nde16} For isoformyl, the convergence with respect to the basis set
is fast and the lowest excited state is well converged with our FCI approach. Hence, we propose a safe TBE of $0.91$ eV for the lowest vertical excitation. Most previous studies did not, once more, discuss
vertical transition energies. However, we are aware of a recent $0.87$ eV CC estimate for the adiabatic energy obtained with a large basis set. \cite{Mor15}

\emph{H$_2$PO and H$_2$PS.}  These two radical homologues of formaldehyde are puckered in their ground state, and CCSDT is already giving very accurate estimates. Indeed, the CCSDT values are consistent with their FCI counterparts,
and one likely needs a triple-$\zeta$ basis set to be close to convergence. The only previous experimental and theoretical studies we are aware of for these two compounds are rather recent. \cite{Gha11,Gri11b,Loo19a} 
They reported: (i) CCSD/{\AVTZ} adiabatic energies of $1.42$ and $3.32$ eV for \ce{H2PO}, \cite{Gha11} and $0.57$ and $2.58$ eV for its sulfur counterpart; \cite{Gri11b} (ii), ROCC3 vertical transitions 
to the lowest $^2A'$ states of $4.35$ eV (\ce{H2PO}) and $2.78$ eV (\ce{H2PS}). \cite{Loo19a} The latter are obviously compatible with the present data.

\emph{Nitromethyl.} For this (comparatively) large derivative, even the UCCSDT/{\AVTZ} calculations are a challenge in terms of computational resources. The calculations converge too slowly with the number of determinants
to ensure valuable FCI extrapolations, except for the second state for which the CCSDT estimate falls within the extrapolation error bar.  Fortunately, for all transitions, the difference between ROCC3 and UCCSDT estimates are small, 
and we can safely propose our CCSDT values as references. These values of $2.05$, $2.38$, $2.56$, and $5.35$ eV do agree rather well with the 2005 ROCC3/Sadlej-TZ estimates of $2.03$, $2.41$, $2.53$ and $5.28$ eV, \cite{Smi05b} 
that remain the most advances carried out previously to the very best of our knowledge. Retrospectively, the MRCI excitation energies of $1.25$ and $1.52$ eV for the two lowest states seem way too low. \cite{Cai94} The measured photoelectron 
spectrum of the related anion indicates the presence of the $^2A_2$ transition at $1.59$ eV in the radical, \cite{Met91} whereas a rough estimate of $4.25$ eV can also be deduced from experimental data for the $^2B_1$ state. 
\cite{Cyr93} We trust that the TBEs given in Table \ref{Table-3} are more trustworthy estimates of the vertical transition energies than these indirect experimental transition energies.

\emph{NO.}  This highly reactive radical is unsurprisingly quite difficult to capture with theoretical approaches and our current TBEs of $6.12$ and $7.21$ eV for the two lowest Rydberg states 
are significantly above the vertical experimental energies  of $5.93$ and $7.03$ eV. \cite{Mau95} Our geometry is associated with a \ce{NO} bond distance of $1.149$ \AA\, slightly larger than the experimental value of $1.115$ \AA. Moreover, 
basis set convergence is slow, so that a quadruple-$\zeta$ basis might still be insufficient to be close to the CBS limit for the second excited state.

\emph{OH.} For \ce{OH}, the convergence of the CC energy with respect to the excitation degree is extremely fast, but the basis set effects are non-negligible. Our TBEs are $4.09$ and $8.11$ eV for the $^2\Sigma^+$ and $^2\Sigma^-$ transitions, respectively. The
former value compares very nicely with the experimental one ($4.08$ eV), \cite{Mau95} and is smaller than previous MRCI estimates of $4.27$ \cite{For91} and $4.22$ eV. \cite{Sza00}  In contrast, for the $^2\Sigma^-$ 
transition, our estimate is  higher than a previously reported value of $7.87$ eV. \cite{For91}

\emph{Vinyl.} For this final radical, we considered four states, two in each spatial symmetry. For the lowest transition of $\pi \rightarrow n$ nature, our FCI/{\AVTZ} result is $3.26\pm0.02$ eV, and one can find 
many previous calculations yielding similar  transition energies: $3.17$ (MRCI), \cite{Wan96b} $3.24$ (MRCI), \cite{Meb97} $3.31$ (CCSD), \cite{Koz06} and $3.30$ eV (CC3), \cite{Loo19a} whereas the measured 0-0 
energy is $2.49$ eV. \cite{Pib99} For the second transition of the same $A''$ symmetry and of $n \rightarrow \pi^\star$ character, the previous theoretical values we are aware of are $4.78$ eV (MRCI) \cite{Meb97} and $4.93$ eV (CCSD). \cite{Koz06}
Our TBE of $4.69$ eV is lower. The lowest $^2A'$ transition is a tricky valence excitation of $\pi \rightarrow \pi^\star$ character with a significant multi-excitation character, 
and we decided to use ROCC for this specific case. It is clear from Table \ref{Table-3} that one needs to go as high as CCSDTQ to be close to FCI.  Our TBE of $5.60$ eV can be compared to previous
estimates of $5.58$ eV (MRCI) \cite{Meb97} or $5.60$ eV (spin-flip CCSD), \cite{Koz06}  which clearly highlights the fantastic accuracy of the spin-flip approach for such transition. Eventually, the last transition of Rydberg character is 
easier to describe at the CC level, with our TBE of $6.20$ eV again close to previously reported results: $6.25$ (MRCI) \cite{Meb97} and $6.31$ eV (CCSD). \cite{Koz06}

\subsubsection{Benchmarks}

\begin{figure}[htp]
  \includegraphics[scale=0.8,viewport=2.8cm 18.3cm 13.3cm 27.5cm,clip]{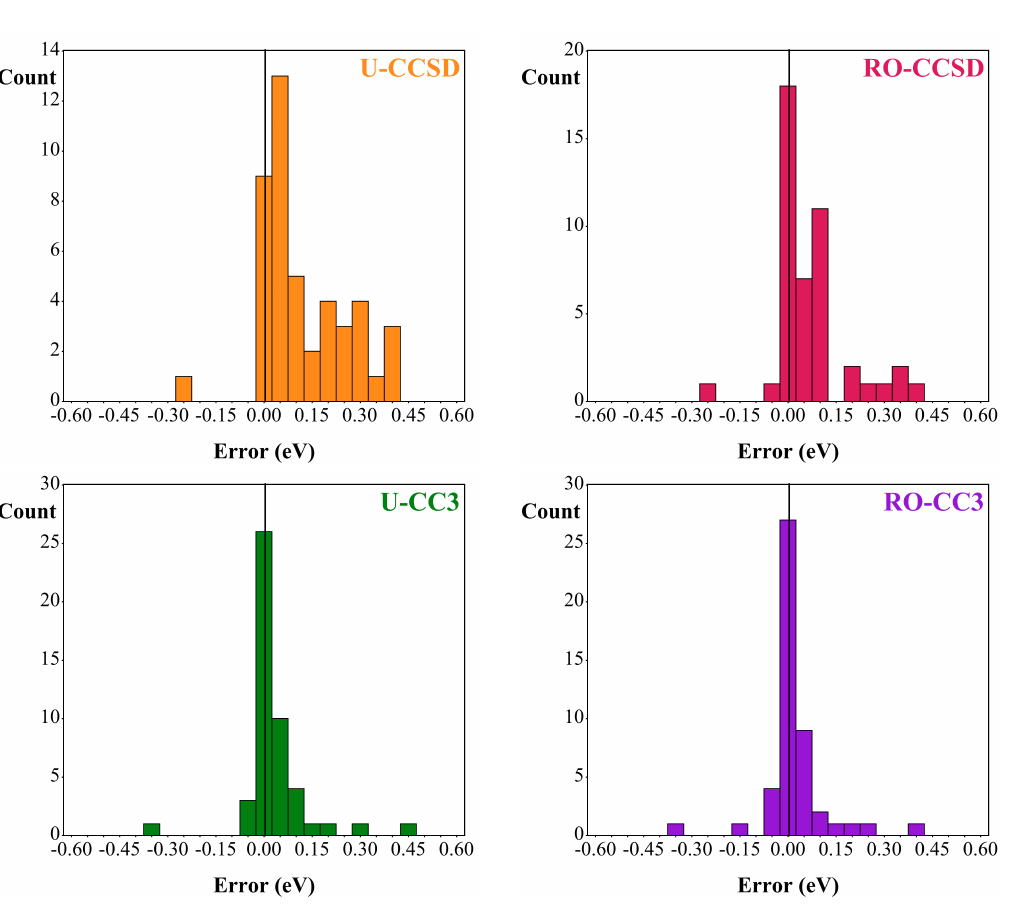}
  \caption{Histograms of the error distribution (in eV) obtained with 4 theoretical methods, choosing the TBE/{\AVTZ} of Table \ref{Table-3} as references (raw data in Table S3). For the CCSD cases, even larger errors (out of scale) are observed.}
  \label{Fig-2}
\end{figure}

As for the exotic set, we have used our TBEs/{\AVTZ} to perform benchmarks of ``lower-order'' methods, and we have especially compared the U and RO versions of CCSD and CC3, considering all transition energies
listed in Table \ref{Table-3} (except three particularly challenging ones that have been omitted, see footnote $g$ in the corresponding Table). The raw data are listed in Table S3 of the SI, whereas Table \ref{Table-4} and Figure \ref{Fig-2} 
gathers the associated statistical data.  As expected from previous works, \cite{Smi05b,Koz06,Loo19a} the excitation energy errors associated with these doublet-doublet transitions in open-shell molecules tend to be larger than for closed-shell systems. Indeed, we note that (i) CCSD overshoots by more than 1 eV the 
transition energies of the second and third excited states of CH; (ii) the MAE obtained with CC3 is $0.05$ eV, five times larger than in the exotic set; and (iii) the error dispersion is obviously larger in Figure \ref{Fig-2} 
than in Figure \ref{Fig-1}. This confirms that accurately describing doublet-doublet transition energies is very challenging. On a more positive note, we observe that the statistical results are improved by using a RO starting 
point instead of the usual U approximation, an effect particularly significant at the CCSD level.

\renewcommand*{\arraystretch}{1.0}
\begin{table}[htp]
\scriptsize
\caption{Statistical values obtained by comparing the results of various methods to the TBE/{\AVTZ} reported in Table S3. See caption of Table \ref{Table-2} for more details.}
\label{Table-4}
\begin{tabular}{lccccc}
\hline
Method 		& Count 	& MSE 	&MAE 	&RMSE 	&SDE 		\\
\hline
UCCSD		&48	&	0.19		&0.20	&0.35	&0.30	\\
ROCCSD		&48	&	0.14		&0.15	&0.30	&0.27	\\
UCC3		&48	&	0.03		&0.06	&0.11	&0.11	\\
ROCC3		&48 	&	0.02		&0.05	&0.10	&0.10	\\
\hline
 \end{tabular}
 \end{table}

\section{Conclusions}

In order to complete our three previous sets of highly-accurate excitation energies, \cite{Loo18b,Loo19b,Loo20a} we have reported here two additional sets of TBEs for: (i) 30 excited states in 
a series of ``exotic'' closed-shell compounds including (at least) one of the following atoms: \ce{F}, \ce{Cl}, \ce{Si}, or \ce{P}; (ii) 51 doublet-doublet transitions in a series of radicals characterized by
an open-shell electronic configuration.  In all cases, we have reported at least {\AVTZ} estimates, the vast majority being obtained at the FCI level, and we have applied increasingly accurate
CC methods to ascertain these estimates. For most of these transitions, it is very likely that the present TBEs are the most accurate published to date (for a given geometry).

For the former exotic set, these TBEs have been used to assess the performances of fifteen ``lower-order'' wave function approaches, including several CC and ADC variants. Consistently
with our previous works, we found that CC3 is astonishingly accurate with a MAE as small as $0.01$ eV and a SDE of $0.02$ eV, whereas the trends 
for the other methods are similar to the one obtained on more standard organic compounds.  In contrast, for the radical set, even the refined ROCC3 method yields a MAE of $0.05$ eV, and 
a rather large SDE of $0.10$ eV. Likewise, the excitation energies obtained with CCSD are much less
satisfying for open-shell derivatives (MAE of $0.20$ eV with UCCSD and $0.15$ eV with ROCCSD) than for the closed-shell systems (MAE of $0.07$ eV).

We hope that these two new sets, which provide a fair ground for the assessments of high-level excited-state models, will be an additional valuable asset for the electronic structure community, and will 
stimulate further developments in the field.

\section*{Acknowledgements}
PFL thanks the \textit{Centre National de la Recherche Scientifique} for funding. This research used resources of (i) the GENCI-TGCC (Grant No.~2019-A0060801738); 
(ii) CALMIP under allocation 2020-18005 (Toulouse); (iii) CCIPL (\emph{Centre de Calcul Intensif des Pays de Loire}); (iv) a local Troy cluster and (v) HPC resources from ArronaxPlus 
(grant ANR-11-EQPX-0004 funded by the French National Agency for Research). 

\section*{Supporting Information Available}
The Supporting Information is available free of charge at https://pubs.acs.org/doi/10.1021/{doi}.

Basis set effects at CC3 level for the exotic set.  Benchmark data.  Multi-reference values for CON. Cartesian coordinates.

\bibliography{biblio-new}

\end{document}